\documentclass{revtex4}
\usepackage{graphicx}
\usepackage{amsmath}
\usepackage{amsfonts}

\begin{document}

\title{Connection between some nonperturbative approaches in QCD}
\author{V. Dzhunushaliev\footnote{Senior Associate of the Abdus Salam ICTP}}
\email{dzhun@krsu.edu.kg}
\affiliation{Dept. Phys.
and Microel. Engineer., Kyrgyz-Russian
Slavic University, Bishkek, Kievskaya Str. 44, 720021, Kyrgyz
Republic}

\begin{abstract}
The connection between two nonperturbative approaches in quantum chromodynamics is considered. The first one is based on a collective coordinate method, the second one on a spin-charge separation. It is shown that both approaches have some close connection: the existence of two condensates which are necessary to confinement. 
\end{abstract}


\maketitle

\section{Introduction}

In this paper we would like to ascertain the connection between some nonperturbative calculations in quantum chromodynamics.  
\par 
The approach (presented in Ref. \cite{niemi1}) is based on the idea that the off-diagonal components of the non-abelian gluon field become composite particles, with a separation between their color-charge and spin degrees of freedom \cite{ludvig1} (see also \cite{ludvig2}-\cite{Faddeev:2006sw}). The authors apply the mean-field approach where a mean-field state is constructed integrating over the charge neutral spin degree of freedom of the off-diagonal gluon. 
\par 
The focus of the second approach \cite{Dzhunushaliev:2006di} is the breaking down of the non-Abelian gauge group into smaller pieces. For example: 
$SU(2) \rightarrow U(1) + coset$ or $SU(3) \rightarrow SU(2) + coset$ or 
$SU(3) \rightarrow U(1) + coset$. The procedure also uses some aspects of an old method by Heisenberg to calculate the n-point Green's function of a strongly interacting, non-linear theory. Using these ideas one can give approximate calculations of the 2 and 4-points Green's function of the theories considered. This method can be called as \textit{the method of collective coordinates} since in this approach some set of quantum degrees of freedom fluctuate in phase.

\section{The basic ideas of the collective coordinates method}
\label{collective}

In this section we will describe in shorten the method of collective coordinates method. The classical SU(N) $(N = 1,2, \cdots , N)$ Yang-Mills equations are 
\begin{equation}
    \partial_\nu F^{B\mu\nu} = 0
\label{sec2-1-10}
\end{equation}
where $F^B_{\mu \nu} = \partial_\mu A^B_\nu - 
\partial_\nu A^B_\mu + g f^{BCD} A^C_\mu A^D_\nu$ 
is the field strength; $B,C,D = 1, \ldots ,N$ are the SU(N) color indices; $g$ is the coupling constant; $f^{BCD}$ are the structure constants for the SU(N) gauge group. In quantizing the system given in Eqs. (\ref{sec2-1-10}) - via Heisenberg's method 
\cite{heisenberg} one first replaces the classical fields by field operators
$A^B_{\mu} \rightarrow \widehat{A}^B_\mu$. This yields the following differential equations for the operators
\begin{equation}
    \partial_\nu \widehat {F}^{B\mu\nu} = 0.
\label{sec2-1-20}
\end{equation}
These nonlinear equations for the field operators of the nonlinear quantum fields can be used to determine expectation values for the field operators 
$\widehat {A}^B_\mu$, where 
$\langle \cdots \rangle = \langle Q | \cdots | Q \rangle$ and $| Q \rangle$ is some quantum state). One can also use these equations to determine the expectation values of operators
that are built up from the fundamental operators $\widehat {A}^B_\mu$. The simple gauge field expectation values, $\langle A_\mu (x) \rangle$, are  obtained by average Eq. \eqref{sec2-1-20} over some quantum state $| Q \rangle$
\begin{equation}
  \left\langle Q \left|
  \partial_\nu \widehat F^{B\mu\nu}
  \right| Q \right\rangle = 0.
\label{sec2-1-30}
\end{equation}
One problem in using these equations to obtain expectation values like $\langle A^B_\mu \rangle$, is that these equations involve not only powers or derivatives of $\langle A^B_\mu \rangle$ ({\it i.e.} terms like $\partial_\alpha \langle
A^B_\mu \rangle$ or $\partial_\alpha 
\partial_\beta \langle A^B_\mu \rangle$) but also contain terms like $\mathcal{G}^{BC}_{\mu\nu} = \langle A^B_\mu A^C_\nu \rangle$. Starting with Eq. \eqref{sec2-1-30} one can generate an operator differential equation for the product $\widehat A^B_\mu \widehat A^C_\nu$ thus allowing the determination of the Green's function $\mathcal{G}^{BC}_{\mu\nu}$
\begin{equation}
  \left\langle Q \left|
  \widehat A^B(x) \partial_{y\nu} \widehat F^{B\mu\nu}(y)
  \right| Q \right\rangle = 0.
\label{sec2-1-40}
\end{equation}
However this equation will in turn contain other, higher order Green's functions. Repeating these steps leads to an infinite set of equations connecting Green's functions of ever increasing order. In fact these equations are the Dyson-Schwinger equatons but ordinary the designation ``Dyson-Schwinger equatons '' is reserved for the application in perturbative quantum field theory. We consider these equations in nonperturbative quantum field theory. 
This construction, leading to an infinite set of coupled, differential equations, does not have an exact, analytical solution and so must be handled using some approximation.
\par
The quantization of equations \eqref{sec2-1-30}-\eqref{sec2-1-40} evidently is very hard and
deriving exact results is probably impossible. In order to do some calculations we give an approximate method which leads to the 2 and 4-points Green's functions only. In order to derive the equations describing the quantized field we average the Lagrangian over a quantum state $\left.\left. \right| Q \right\rangle$
\begin{equation}
\begin{split}
    \left\langle Q \left| \widehat {\mathcal L}_{SU(N)}(x) \right| Q \right\rangle =
    &\left\langle \widehat {\mathcal L}_{SU(N)} \right\rangle = \\
    &\frac{1}{2}
    \left\langle
      \left( \partial_\mu \widehat A^B_\nu (x) \right)
      \left( \partial^\mu \widehat A^{B\nu} (x) \right) -
      \left( \partial_\mu \widehat A^B_\nu (x) \right)
      \left( \partial^\nu \widehat A^{B\mu} (x) \right)
    \right\rangle + \\
    &\frac{1}{2}    g f^{BCD}
    \left\langle
      \left( \partial_\mu \widehat A^B_\nu (x)-
      \partial_\nu \widehat A^B_\mu (x)\right)
      \widehat A^{C \mu} (x)\widehat A^{D \nu}(x)
    \right\rangle + \\
    &\frac{1}{4}g^2 f^{BC_1D_1} f^{BC_2D_2}
    \left\langle
      \widehat A^{C_1}_\mu (x)\widehat A^{D_1}_\nu (x)
      \widehat A^{C_2 \mu} (x)\widehat A^{D_2\nu} (x)
    \right\rangle .
\end{split}
\label{sec3-5}
\end{equation}
Now we will detail the kind of physical situations we wish to describe. The model given here is similar to stationary turbulence when there are time dependent fluctuations in any point of the liquid but all averaged quantities are time independent. For a QFT this means that all Green's functions are time independent and there is a correlation between quantum fields in different points at one moment
\begin{eqnarray}
    \left\langle
    A^{B_1}_{\mu_1}(x^\alpha_1) \cdots A^{B_n}_{\mu_n}(x^\alpha_2)
    \right\rangle &\neq &0 ,
\label{sec3-10}\\
  t_1 &=&  \cdots = t_n ,
\label{sec3-20}\\
  \vec{r}_1 &\neq& \cdots \neq \vec{r}_n .
\label{sec3-30}
\end{eqnarray}
In linear and perturbative QFT this is not the case because the interaction is carried by quanta which move with a speed less than or equal to the speed of light. In these theories the correlation between quantum fields in different points at the same time is zero.
\begin{eqnarray}
    \left\langle
    A^{B_1}_{\mu_1}(x^\alpha_1) \cdots A^{B_n}_{\mu_n}(x^\alpha_2)
    \right\rangle &= &0 ,
\label{sec3-40}\\
  t_1 &=&  \cdots = t_n ,
\label{sec3-50}\\
  \vec{r}_1 &\neq& \cdots \neq \vec{r}_n .
\label{sec3-60}
\end{eqnarray}
In this sense one can say that nonperturbative QFT in some physical situations is very close to turbulence, {\it i.e.} in nonperturbative QFT there may exist extended objects where quantized fields at all points are correlated between themselves (example from QCD are flux tubes and glueballs). Such objects fall into two categories
\begin{enumerate}
\item
The averaged value of all quantized fields are zero (all components are in one or more  collective modes)
    \begin{equation}
        \left\langle
        A^B_\mu(x)
        \right\rangle = 0 .
    \label{sec3-70}
    \end{equation}
    But the square of these fields are nonzero
    \begin{equation}
        \left\langle
        \left( A^B_\mu(x) \right)^2
        \right\rangle \neq 0
    \label{sec3-80}
    \end{equation}
    \item
    The averaged value of some components of the quantized fields are nonzero (approximately they can be considered on the classical level) and the averaged value of some zero (they are in one or more collective modes), and the square of some components is nonzero
    \begin{eqnarray}
        \left\langle
        A^{B_1}_\mu(x)
        \right\rangle &\neq & 0
        \quad \text{for some} \quad {B_1} \in 1,2, \cdots , N
    \label{sec3-90}\\
    \left\langle
        A^{B_2}_\mu(x)
        \right\rangle &=& 0
        \quad \text{but } 
        \left\langle
        \left( A^{B_2}_\mu(x) \right)^2
        \right\rangle \neq 0 \quad 
        \text{for remaining} \quad {B_2} \in 1,2, \cdots , N
    \end{eqnarray}
    The most natural case is when $A^{B_1}_\mu$ belongs to a small subgroup
    of $G = SU(N)$ gauge group (for example, to $U(1) \subset SU(2)$ or $SU(2) \subset SU(3)$ or $U(1) \times SU(2) \subset SU(3)$ all these cases will be considered below) and $A^{B_2}_\mu$ are the coset components $SU(N) / G$. 
\end{enumerate}
In the first case the quantized fields are in a completely disordered phase. In the second case one has both ordered and disordered  phases. 

\section{The concrete realization of the collective coordinates method}

The key idea in this approach is \textit{to cut off an infinite equations set connecting all Green functions.} For this we average the Lagrangian of a gauge theory 
\begin{equation}
  \left\langle \mathcal L \right\rangle = 
  \left\langle F^A_{\mu \nu} F^{A\mu \nu}
  \right\rangle.
\label{4-10}
\end{equation}
Schematically the averaged Lagrangian has 
$\left\langle (A)^2 \right\rangle, \left\langle (A)^3 \right\rangle$ and 
$\left\langle (A)^4 \right\rangle$ terms. We assume that there are components of gauge potential $A^a_\mu$
\begin{equation}
  \left\langle A^a_\mu \right\rangle \approx A^a_\mu
\label{4-20}
\end{equation}
where $A^a_\mu \in G \subset SU(N)$ is a small subgroup. In this case we will say that $A^a_\mu$ is in an ordered phase. As well there are components of gauge potential $A^m_\mu$
\begin{equation}
  \left\langle A^m_\mu \right\rangle = 0 \quad \text{but }
  \left\langle \left( A^m_\mu \right)^2 \right\rangle \neq 0.
\label{4-25}
\end{equation}
In this case we will say that $A^m_\mu$ is in a disordered phase. 
\par 
The main problem in this approach is calculating two and four points Green functions 
\begin{eqnarray}
  \mathcal G^{BC}_{\mu \nu} (x,y) &= & \left\langle 
  	A^B_\mu (x) A^C_\nu (y)
  \right\rangle ,
\label{4-30}\\
  \mathcal G^{BCDE}_{\mu \nu \alpha \beta} (x,y, z, u) &= & \left\langle 
  	A^B_\mu (x) A^C_\nu (y) A^D_\alpha (z) A^E_\beta (u)
  \right\rangle .
\label{4-40}
\end{eqnarray}
Both functions $\mathcal G$ are non-local. The different approximation for $\mathcal G$ will result in the different non-perturbative approximate approaches. For example, one can to single out with various ways the color and spacetime indices. 

\subsection{The collective coordinates method for the SU(2) gauge group: the Ginzburg-Landau equation}

The application of this method to SU(2) gauge group (where $A^3_\mu \in U(1) \subset SU(2)$ is in ordered phase and $A^{1,2}_\mu \in SU(2)/U(1)$) gives us \cite{Dzhunushaliev:2002xr} 
\begin{equation}
  \mathcal G^{mn}_{\mu \nu} (x,y) = \left\langle A^m_\mu (x) A^n_\nu (y)
  \right\rangle = 
  - \frac{1}{5} \eta_{\mu \nu} \delta^{mn} \phi^*(x) \phi(y)
\label{4a-10}
\end{equation}
where $m,n = 1,2$ lead to 
\begin{equation}
  \left \langle {\mathcal L} \right\rangle = 
	- \frac{1}{4}f_{\mu \nu} f^{\mu \nu} + 
    \left( D_\mu \phi^* \right) \left( D^\mu \phi \right) - 
   \frac{6 g^2}{25} \left| \phi \right|^4 
\label{sec4a-20}
\end{equation}
where $f_{\mu \nu} = \partial_\mu A^3_\nu - \partial_\nu A^3_\mu$ can be considered as the ordinary electromagnetic field. As well here we assume that 
\begin{equation}
\begin{split}
  \left\langle
  A^m_\alpha(x) A^n_\beta(y) A^p_\mu(z) A^q_\nu(u) 
  \right\rangle = &
  \delta^{mp} \delta^{nq} \eta_{\alpha\mu} \eta_{\beta\nu} 
  \mathcal{G}(x,z) \mathcal{G}(y,u) + 
  \delta^{mq} \delta^{np} \eta_{\alpha\nu} \eta_{\beta\mu} 
  \mathcal{G}(x,u) \mathcal{G}(y,z) + \\ 
  &\delta^{mn} \delta^{pq} \eta_{\alpha\beta} \eta_{\mu\nu} 
  \mathcal{G}(x,y) \mathcal{G}(z,u) 
\label{4a-40}
\end{split}
\end{equation}
In Ref. \cite{Dzhunushaliev:2002xr} it is shown that a tachyonic mass term can be generated for the off-diagonal gauge fields of a pure SU(2) Yang-Mills via a condensation of ghost and anti-ghost fields. In this case 
\begin{equation}
  \left \langle {\mathcal L} \right\rangle = 
	- \frac{1}{4}f_{\mu \nu} f^{\mu \nu} + 
    \left( D_\mu \phi^* \right) \left( D^\mu \phi \right) - 
   \frac{6 g^2}{25} \left| \phi \right|^4 + 
   \frac{v g^2}{20 \pi} | \varphi |^2
\label{sec4a-50}
\end{equation}
Let us note the following essential thing: the approxiamtion \eqref{4a-10} splits the non-local function $\mathcal G^{mn}_{\mu \nu} (x,y)$ into the product of two local functions $\phi^*(x)$ and $\phi(y)$.

\subsection{The collective coordinates method for the SU(3) gauge group: flux tube}

In this case \cite{Dzhunushaliev:2006di} the initial degrees of freedom 
$A^B_\mu, B=1, \cdots , 8$ are decomposed on two sets: (a) almost classical degrees of freedom $A^a_\mu \in SU(2) \subset SU(3), a=1,2,3$ and (b) pure quantum degrees of freedom 
$A^m_\mu \in SU(3)/SU(2) , m=4,5,6,7,8$. The components $A^m_\mu$ are in a disordered phase and form a condensate of the gluon filed. The Green function 
\begin{equation}
  \mathcal G^{mn}_{\mu \nu} (x,y) \approx \left\langle A^m_\mu (x) A^n_\nu (y)
  \right\rangle = 
  - \frac{1}{3} \eta_{\mu \nu} f^{mpb} f^{npc} \phi^{b}(x) \phi^c(y)
\label{4b-10}
\end{equation}
where $f^{abc}$ are the structural constants of the SU(3) gauge group $m,n = 4,5,6,7,8$ and 
$b,c = 1,2,3$. In Ref. \cite{Dzhunushaliev:2006di} the scalar field $\phi^b$ is real function but it is not too hard to generalize to the complex scalar field. Similar to \eqref{4a-10} we have the decomposition of a non-local function $\mathcal G^{mn}_{\mu \nu} (x,y)$ into some linear combination of local functions $\phi(x)$. Let us note that the components $A^m_\mu$ with the different spacetime indices $\mu, \nu$ do not interact each with other but the components $A^m_\mu$ with the different color indices $m,n$ interact each with other. 
\par 
The four point Green function can be decomposed in a similar manner with Eq. \eqref{4a-40}
\begin{equation}
\begin{split}
  \left\langle
  A^m_\alpha (x) A^n_\beta (y) A^p_\mu (z) A^q_\nu (u)
  \right\rangle \approx & \frac{1}{3}
  \left(
  \delta^{mn}\delta^{pq} \eta_{\alpha\beta} \eta_{\mu\nu} +
  \delta^{mp}\delta^{nq} \eta_{\alpha\mu} \eta_{\beta\nu} +
  \delta^{mq}\delta^{np} \eta_{\alpha\nu} \eta_{\beta\mu}
  \right)  \\ 
  &\left(
    \delta_{ab} \delta_{cd} + \delta_{ac} \delta_{bd} +
    \delta_{ad} \delta_{bc}
  \right) 
  \phi^a (x) \phi^b(y) \phi^c (z) \phi^d(u) 
\label{4b-15}
\end{split}
\end{equation} 
After some assumptions and simplifications one can receive 
\begin{equation}
\begin{split}
   \left\langle \mathcal{L}_{SU(3)} \right\rangle \approx	&
  - \frac{1}{4}  h^a_{\mu\nu} h^{a\mu\nu} +
  \frac{1}{2}   \left(
  \partial_\mu \phi^a + g \epsilon^{abc} a^b_\mu \phi^c
  \right)^2 
  - \frac{\lambda}{4} \left( \phi^a (x) \phi^a(x) \right)^2 
  \\
  &+ 
  \frac{g^2}{2} a_{\mu} ^b \phi ^b a^{c \mu} \phi ^c 
  + \frac{m^2_\phi}{2} (\phi ^a \phi ^a ) + 
  \left( m^2 \right)^{ab} a^a_\mu a^{b\mu}
\end{split}
\label{4b-20}
\end{equation}
here is assumed that there are mass terms $\left( m^2 \right)^{ab} a^a_\mu a^{b\mu}$ (which breaks the SU(2) gauge invariance of the initial Lagrangian) and $\frac{m^2_\phi}{2} (\phi ^a \phi ^a )$. These terms can be derived if to change the decomposition \eqref{4b-15} from 
$\left\langle A^4 \right\rangle = \left\langle A^2 \right\rangle^2$ to 
$\left\langle A^4 \right\rangle = \left\langle A^2 \right\rangle^2 + 
c_1 \left\langle A^2 \right\rangle + c_2$ where $c_{1,2}$ are some constants. 
\par
\begin{figure}[h]
  \begin{center}
	\fbox{ 
	\includegraphics[totalheight=7cm,width=7cm,origin=c]{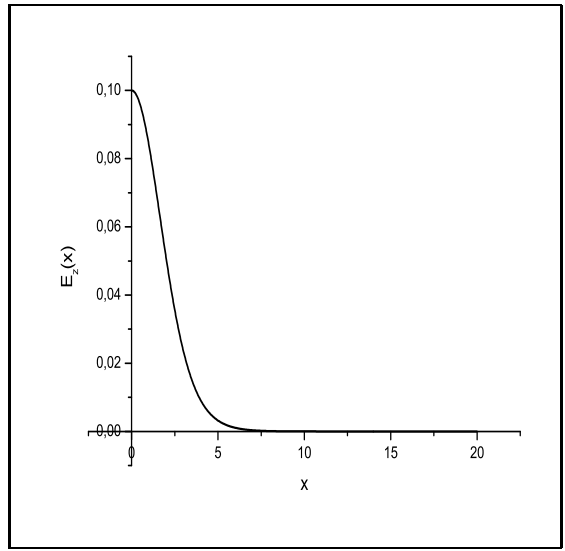}}
  \caption{The longitudinal electric field.}
  \label{energy}
  \end{center}
\end{figure}

 
The numerical solution of the corresponding field equations \cite{Dzhunushaliev:2006di} shows that there is a flux tube solution filled with the color longitudinal electric field $E^3_z$ and the linear energy density presented in Fig. \ref{energy}. 

\subsection{The collective coordinates method for the SU(3) gauge group: the bag of quantum fields}
\label{bag}

In this case all degrees of freedom are pure quantum but there are two kinds of collective modes. For two point Green function 
\begin{equation}
	\left\langle 
	  \widehat A^B_\alpha (x) \widehat A^C_\beta (y) 
	\right\rangle \approx 
	-\eta_{\alpha \beta} f^{ACD} f^{BCE} \phi^{D} (x) \phi^E(y)
\label{4c-10}	
\end{equation}
here the same words about real and complex scalar fields can be said as in the previous section. Four point Green function is decomposed similar to previous section 
\begin{equation}
\begin{split}
	&\left\langle 
	  \widehat A^B_\alpha (x) \widehat A^C_\beta (y)
	  \widehat A^D_\gamma (z) \widehat A^R_\delta (u)
	\right\rangle \approx 
	\left\langle 
	  \widehat A^B_\alpha (x) \widehat A^C_\beta (y)
	\right\rangle  
	\left\langle 
	  \widehat A^D_\gamma (z) \widehat A^R_\delta (u)
	\right\rangle + \\
	&\left\langle 
	  \widehat A^B_\alpha (x) \widehat A^D_\gamma (z)
	\right\rangle  
	\left\langle 
	  \widehat A^C_\beta (y) \widehat A^R_\delta (u)
	\right\rangle + 	
	\left\langle 
	  \widehat A^B_\alpha (x) \widehat A^R_\gamma (u)
	\right\rangle  
	\left\langle 
	  \widehat A^C_\beta (y) \widehat A^D_\gamma (z)
	\right\rangle .
\end{split}	
\label{4c-20}
\end{equation}
For the concrete calculations of \eqref{4c-20} it is very essential the assumtions that for this case the color space is \textit{anisotropic}. Roughly speaking, it means that 
\begin{eqnarray}
  \left\langle 
	  \widehat A^B_\alpha (x) \widehat A^C_\beta (y)
	  \widehat A^D_\gamma (z) \widehat A^E_\delta (u)
	\right\rangle &= & \lambda_1 f^{BCDE}_{\alpha \beta \gamma \delta} (x,y,z,u) 
	\quad \text{for} \quad B,C,D,E = 1,2,3 ;
\label{4c-30}\\
  \left\langle 
	  \widehat A^B_\alpha (x) \widehat A^C_\beta (y)
	  \widehat A^D_\gamma (z) \widehat A^E_\delta (u)
	\right\rangle &= & \lambda_2 f^{BCDE}_{\alpha \beta \gamma \delta} (x,y,z,u) 
	\quad \text{for} \quad B,C,D,E = 4,5,6,7,8 ;
\label{4c-40}\\
  \left\langle 
	  \widehat A^B_\alpha (x) \widehat A^C_\beta (y)
	  \widehat A^D_\gamma (z) \widehat A^E_\delta (u)
	\right\rangle &= & f^{BCDE}_{\alpha \beta \gamma \delta} (x,y,z,u) 
	\quad \text{if} 
	\left\{	
		\begin{array}{ll}
			\text{some indices} & B,C,D,E = 1,2,3\\
			\text{some indices} & B,C,D,E = 4,5,6,7,8\\
		\end{array}
	\right.
\label{4c-50}
\end{eqnarray}
The averaged Lagrangian is now 
\begin{equation}
\begin{split}
	\frac{g^2}{4} \mathcal {L}_{eff} = &- \frac{1}{2}\left( \partial_\mu \phi^A \right)^2 + 
	\frac{\lambda_1}{4} 
	\left[ \phi^a \phi^a - \mu_1^2 
	\right]^2 + \\
	&\frac{\lambda_2}{4} 
	\left[ \phi^m \phi^m - \mu_2^2 
	\right]^2 - \frac{\lambda_2}{4} \mu_2^4 + 
	\left( \phi^a \phi^a \right) \left( \phi^m \phi^m \right) 
\end{split}
\label{4c-60}
\end{equation}
The quantum Lagrangian describes the interaction of two scalar fields $\phi^a$ and $\phi^m$ which present two and four Green functions of SU(3) gauge fields. The numerical calculations of the corresponding field equations show that exists a spherically symmetric solution which can be interpret as a bag filled with the quantum SU(3) gauge fields \cite{Dzhunushaliev:2006di}. The solution has finite energy for some value of $\mu_{1,2}=\mu^*_{1,2}$ only, in other words mathematically the search of a regular solution is nonlinear eigenvalue problem for eigenfunctions $\phi^{a,m}$ and eigenvalues $\mu_{1,2}$. 

\subsection{The collective coordinate method for the SU(3) gauge group: confinement of the field angular momentum}
\label{momentum}

In this case in contrast with the previous subsection there is almost classical degree of freedom $A^8_\mu$ \cite{Dzhunushaliev:2006di} and other components of the SU(3) gauge potential are pure quantum degrees of freedom similar to the previous section with one exception: the index $m=4,5,6,7$ in Eq. \eqref{4b-10}. The averaged Lagrangian is now 
\begin{equation}
\begin{split}
    \left\langle \mathcal{L}_{SU(3)} \right\rangle = &
    - \frac{1}{4} h_{\mu \nu} h^{\mu \nu} +
    \frac{1}{2} \left( \partial_\mu \phi^a \right)
    \left( \partial^\mu \phi^a \right) +
    \frac{1}{2} \left( \partial_\mu \phi^m \right)
    \left( \partial^\mu \phi^m \right) - \\
    &
    \frac{\lambda_1}{4} \left[
        \left( \phi^a \phi^a \right) - \mu_1^2
    \right]^2 -
    \frac{\lambda_2}{4} \left[
        \left( \phi^m \phi^m \right) - \mu_2^2
    \right]^2 - \frac{\lambda_2}{4} \mu_2^4 - \\
    &
    \frac{k_2}{6} \left( \phi^a \phi^a \right) \left( \phi^m \phi^m \right) +
    \left( b_\mu b^\mu \right) \phi^a \phi^a -
    \frac{1}{2} \left( m^2 \right)^{\mu \nu} b_\mu b_\nu
\end{split}
\label{4d-10}
\end{equation}
which describes the interaction of the U(1) gauge field $A_\mu$, with two scalar fields $\phi^a$ ($a=1,2,3$) and $\phi^m$ ($m=4,5,6,7$). The solution of the corresponding field equations shows that the ball of quantum fields $A^B_\mu, B = 1,2,3,4,5,6,7$ can confine $A^8_\mu$ component of the SU(3) gauge potential. The space profile of $A^8_\mu$ is made by such a way that a field angular momentum of $A^8_\mu$ appears. 
\par 
It is necessary to note that all solutions presented in this section exist only for some discrete choice of parameters of the corresponding equations. Mathematically it means that we have a non-linear eigenvalue problem and physically that we actually solve a quantum problem. 

\section{A spin-charge separation}
\label{slave}

In this section we would like to briefly outline the non-perturbative approach for QCD following to \cite{Niemi:2005qs} (with all corresponding references).
\par 
The slave-boson decomposition (spin-charge separation) of the $SU(2)$ gauge field $A^a_\mu$ ($a=1,2,3$ and $\mu=0,1,2,3$) proceeds as follows \cite{ludvig1}: We first separate the diagonal 
Cartan component $A^3_\mu = A_\mu$ from the off-diagonal components $A^{1,2}_\mu$, and combine the latter into the complex field $W_\mu = A^1_\mu + i A^2_\mu$. We then 
introduce a complex vector field $\vec{ e}_\mu$ with
\begin{equation}
	\vec{ e}_\mu \vec{ e}_\mu = 0 \qquad \text{and}
	\qquad \vec{ e}_\mu \vec{ e}^*_\mu = 1.
\label{5-10}
\end{equation}
We also introduce two spinless complex scalar fields $\psi_1$ and $\psi_2$. The ensuing decomposition of $W_\mu$ is \cite{ludvig1}
\begin{equation}
	W_\mu = A^1_\mu + i A^2_\mu = \psi_1 \vec{ e}_\mu + 
	\psi_2^* \vec{ e}^*_\mu.
\label{5-20}
\end{equation} 
A decomposition of $W_\mu$ into spinless bosonic scalars $\psi_{1,2}$ which describe the gluonic holons that carry the color charge of the $W_\mu$, and a color-neutral spin-one vector $\vec{ e}_\mu$ which is 
the gluonic spinon that carries the statistical spin degrees of freedom of $W_\mu$. 
\par
The mean-field approximation applied in this case is averaging the $SU(2)$ Yang-Mills action both over the color-spinon $\vec{ e}_\mu$ and the Cartan component $A_\mu$ of the gauge field. Since we are only interested in the phase structure of the ensuing 
mean-field theory, it is sufficient to consider the free energy in a London limit where the slave-boson condensates
\begin{equation}
	\rho_{1,2}^2 = \langle |\psi_{1,2}|^2 \rangle 
\label{5-30}
\end{equation} 
are spatially uniform.
\par 
The integration over $A_\mu$ and $\vec{ e}_\mu$ can be performed in various different ways. 
In Ref. \cite{Niemi:2005qs} the free energy in mean-field approximation is investigated. After some calculations the final version of the free energy is 
\begin{equation}
	F = \frac{1}{2}\left( \rho_1^2 - \rho_2^2 \right)^2 \biggl( 1 + \lambda 
	\ln \left[ \left( \rho_1^2-\rho_2^2 \right)^2 \right] \biggr) \quad.
\label{5-40}
\end{equation}
where $\lambda$ is some constant. 

The generic features of this free energy, a ridge along the lines $\rho_1=\pm\rho_2$, and a narrow hyperbolic valley on both sides of these lines, are independent of $\lambda$, but the depth of the valleys and steepness of the potential are more prominent for larger values of $\lambda$, as used here.
\par 
The four branches of the hyperbola that minimize \eqref{5-40},
\begin{equation}
	(\rho_{1}^{2}-\rho_{2}^{2})^2 = \exp\left(\frac{1-\lambda}
	{\lambda}\right)\quad,
\label{5-50}
\end{equation}
are separated by (non-analytic) ridges along the lines $\rho_{1}=\pm \rho_{2}$. At the minima along the hyperbolic valleys the free energy is given by
\begin{equation}
	F_{\text{min}} =- \frac{1}{2}\exp\left(\frac{1-\lambda}{\lambda}\right)\quad.
\label{5-60}
\end{equation}
This ground state is highly degenerate, but the combination on the left-hand side  of \eqref{5-60} is not the proper gauge invariant condensate. The gauge invariant 
condensate is given by 
\begin{equation}
	\rho^2 = \rho_1^2 + \rho_2^2 
\label{5-70}
\end{equation}
and one can employ it to remove the infinite degeneracy of the hyperbolic vacuum.
\par 
From \eqref{5-60} we conclude that the ground state value $\rho^2 = v^2$ of the gauge invariant condensate \eqref{5-70} is bounded from below by a non-vanishing
quantity,
\begin{equation}
	\rho^2 = \rho_1^2 + \rho_2^2 = v^2  \geq  |\rho_1^2 - \rho_2^2| =
	\exp\left(\frac{1-\lambda}{2\lambda}\right).
\label{5-80}
\end{equation}
When $v^2$ is larger than the lower bound in \eqref{5-80}, there are eight solutions $(\rho_1,\rho_2)$ to the equations that define the vacuum
\begin{eqnarray}
	\rho_1^2 + \rho_2^2 & = & v^2 \quad,\nonumber \\
	\rho_1^2 - \rho_2^2 & = & \pm \exp\left( \frac{1-\lambda}{2\lambda}\right)
	\quad.\label{vaceq1}
\label{5-90}
\end{eqnarray}
But when $v^2$ coincides with the lower bound there are only four solutions,
\begin{eqnarray}
	\rho_1=\pm v \quad&\& &\quad \rho_2=0 
	\quad,\nonumber \\
	\rho_1=0 \quad&\& &\quad \rho_2=\pm v
\label{5-100}
\end{eqnarray}
which correspond to the vertices of the hyperbola. 
\par
The solutions of \eqref{5-100} describe the generic situation where both condensates are non-vanishing. The remaining ground state is doubly degenerate under exchange of $\rho_1$ and $\rho_2$, which correspond to the physical scenario that in general the London
limit densities are unequal.
\par
Finally, the degenerate solutions \eqref{5-100} correspond to the limit where one of the two condensates vanishes, and again by selecting the physical quadrant $\rho_{1,2} \geq 0$ one can remove the degeneracy.
\par 
According to \eqref{5-80} the ground state value of \eqref{5-70} is non-vanishing for all non-vanishing values of the coupling constant $\lambda $. This suggests that in the Yang-Mills theory the gauge invariant condensate is also nonvanishing for all
non-vanishing values of the coupling constant. This would mean that the mass gap in the Yang-Mills theory is present for all values of the coupling, and it vanishes only asymptotically in the short distance limit where the gluons become asymptotically free and massless. 
\par 
The classical treatment of the mean-field theory suggests that the condensate is always non-vanishing, hence a mass gap is present for all nontrivial values of the coupling. One can to inspect what effects spatially homogeneous quantum fluctuations around the classical mean-field value have on this condensate. For this it is necessary to improve the free energy so that it also includes the contribution from the momenta $\pi_{1,2}$ that are canonically conjugate to the (spatially homogeneous) condensates $\rho_{1,2}$. 
\par 
The result of Ref. \cite{Niemi:2005qs} suggests that the possibility of a spin-charge separation in the Yang-Mills theory may occur. Furthermore, there is a need to address theoretical issues such as electric-magnetic duality and the description of the Yang-Mills theory in
terms of the spin-charge separated dual variables. 

\section{The comparison of the collective coordinate and spin-charge separation methods}

It is very important to compare some different approaches in a non-perturbative quantum field theory. Every approach has own weakness and advantages. If some conclusions in these approaches coincide then it gives us some confidence that both approaches are correct.
In this section we would like to compare the collective coordinate method briefly sketched in section \ref{collective} with a spin-charge separation briefly sketched in section  \ref{slave}. 
\par 
The deep connection between the collective coordinate method (CCM) and the spin-charge separation method (SCSM) is the existence of two condensates $\phi^a$ and $\phi^m$ in Eq's \eqref{4c-60} \eqref{4d-10} in CCM and $\rho_1$ and $\rho_2$ in Eq. \eqref{5-30} SCSM. The field equations for the Lagrangians \eqref{4c-60}, \eqref{4d-10} have regular solutions (with a finite energy) for two \textit{different} scalar fields $\phi^{a,m}$ only. SBDM tells us similar thing: the condensates $\rho_1$ and $\rho_2$ should be different in order to have a mass gap Eq's \eqref{5-90} \eqref{5-100}. 
\par
\begin{figure}[h]
\begin{minipage}[t]{.45\linewidth}
  \begin{center}
  \includegraphics[totalheight=7cm,width=7cm,origin=c,angle=270]
  {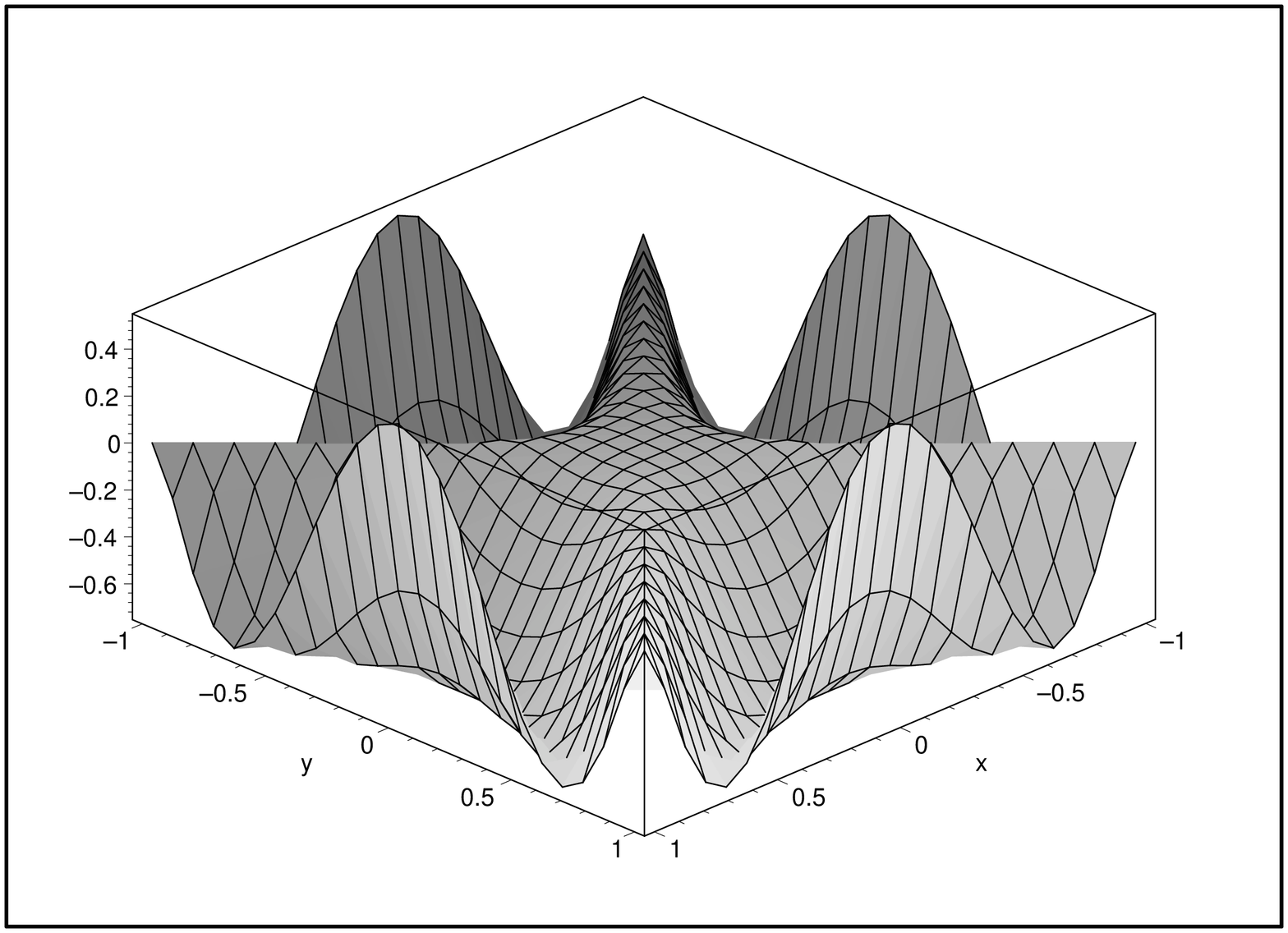}
  \caption{}
  \label{free}
  \end{center}
\end{minipage}\hfill
\begin{minipage}[t]{.45\linewidth}
 \begin{center}
  \includegraphics[totalheight=7cm,width=7cm,origin=c,angle=270]
  {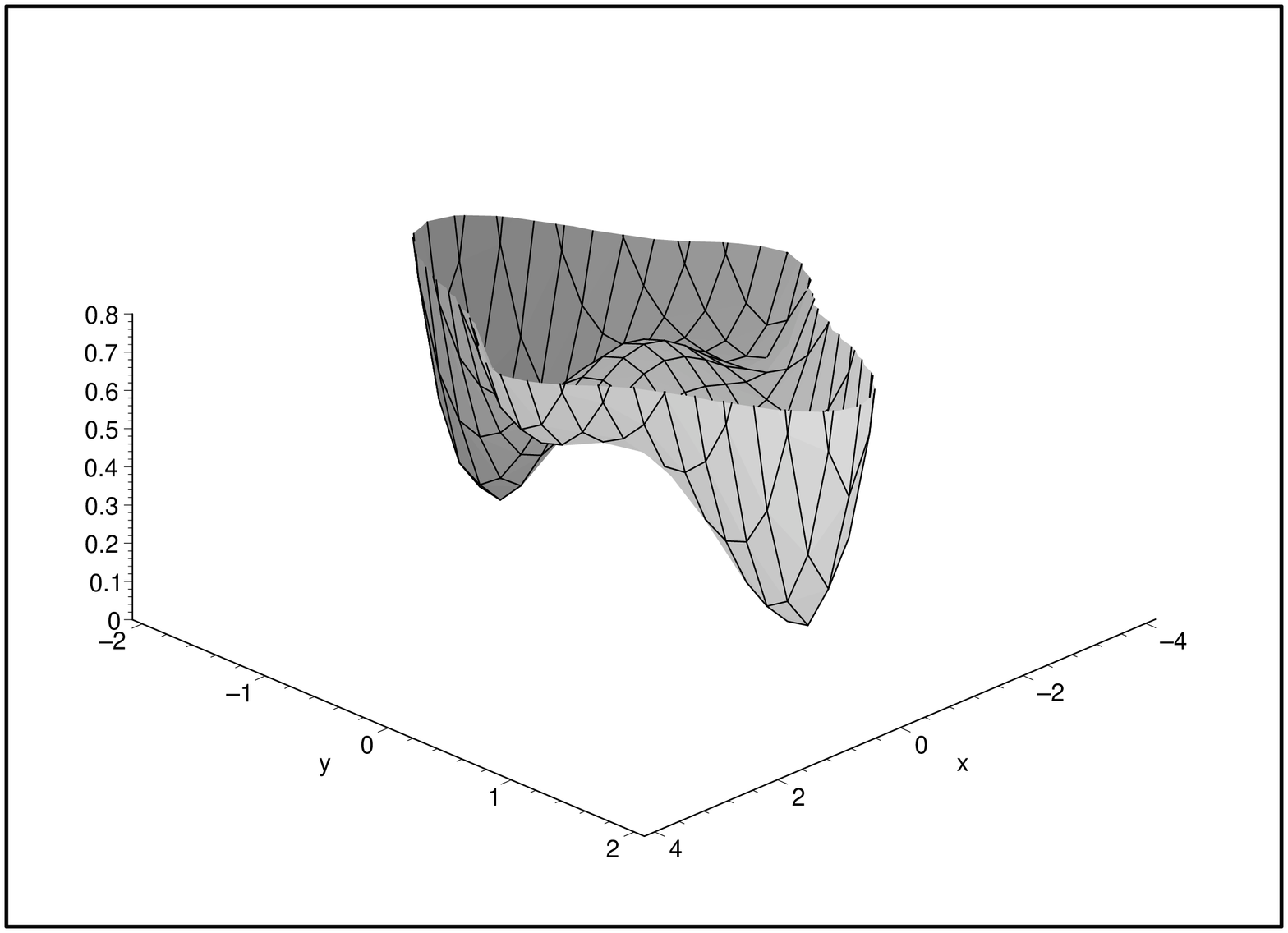}
  \caption{}
  \label{potential}
  \end{center}
\end{minipage}\hfill
\end{figure}

 
In Ref. \cite{Niemi:2005qs} the free energy \eqref{5-40} is investigated. $F$ can be connected with the energy by the following way 
\begin{equation}
  F = V - TS
\label{7-10}
\end{equation}
where $V,T,S$ are correspondingly the energy, temperature and entropy. In Fig. \ref{free} the free energy of SCSM and the potential energy of CCM are plotted. It is visible that both pictures are a little similar. $F$ has two highly degenerated local maxima on the lines 
\begin{equation}
  \rho_1^2 - \rho_2^2 = 0
\label{7-20}
\end{equation}
and four global degenerated minima located on the hyperbola 
\begin{equation}
  \rho_1^2 - \rho_2^2 = \mathrm{const}
\label{7-30}
\end{equation}

The potential (see. Fig. \ref{potential})
\begin{equation}
  V\left( \phi^a, \phi^m \right) = \frac{\lambda_1}{4} \left[
    \left( \phi^a \phi^a \right) - \mu_1^2 \right]^2 -
    \frac{\lambda_2}{4} \left[
        \left( \phi^m \phi^m \right) - \mu_2^2
    \right]^2 - \frac{\lambda_2}{4} \mu_2^4
\label{7-40}
\end{equation}
has one local non-degenerate maximum $\phi^a = \phi^m = 0$ instead of \eqref{7-20} and two local non-degenerate and two global non-degenerate minima $\phi^a = \mu_1$ $\phi^m = 0$  instead of \eqref{7-30}. This difference probably is connected with the fact that the free energy $F$ has derived from one-loop approximation but the potential $V$ is derived by the assumption about a non-perturbative structure of two-point Green function. 
\par
\begin{figure}[h]
\begin{minipage}[t]{.45\linewidth}
  \begin{center}
  \includegraphics[height=7cm,width=7cm]{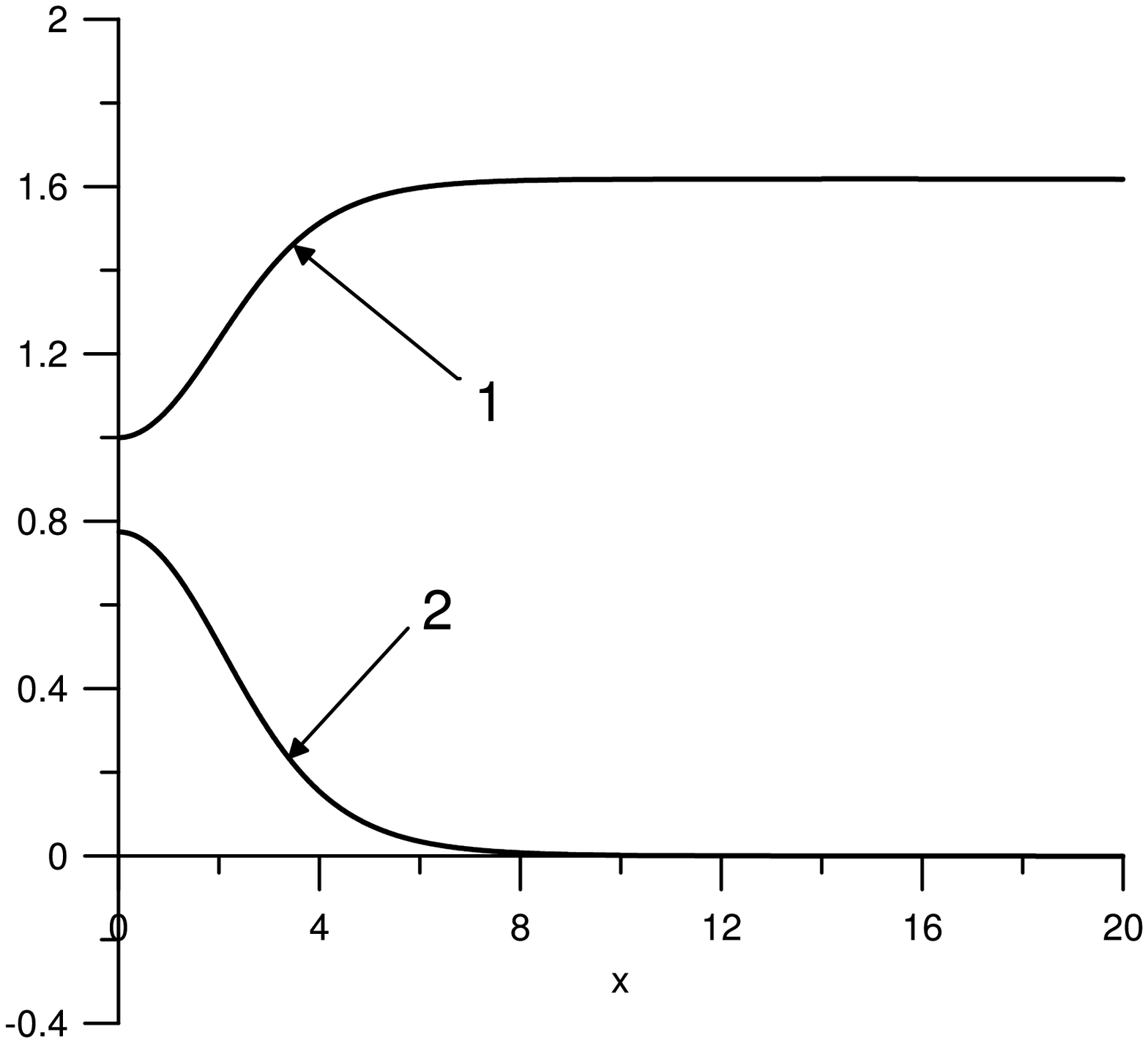}
  \caption{1 - the condensate $\phi^a(r)$, 2 - the condensate $\phi^m(r)$.}
  \label{condensates}
  \end{center}
\end{minipage}\hfill
\begin{minipage}[t]{.45\linewidth}
 \begin{center}
  \includegraphics[height=7cm,width=7cm]{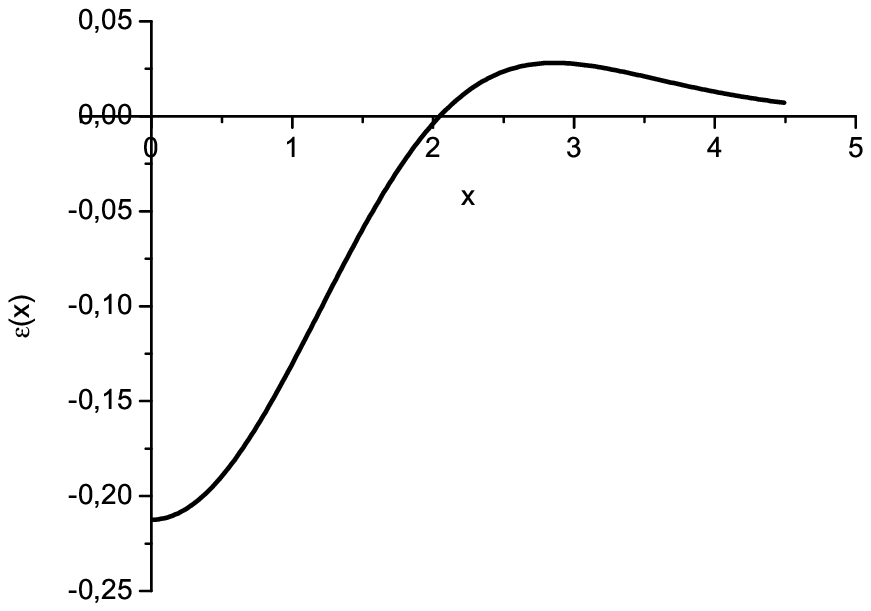}
  \caption{The energy density for the bag of quantum SU(3) gauge field.}
  \label{energy_glueball}
  \end{center}
\end{minipage}\hfill
\end{figure}

 
The spherically symmetric solution from the section \ref{bag} (a bag filled with quantum SU(3) gauge field which are presented by the scalar fields $\phi^a$ and $\phi^m$) is plotted in Fig. \ref{condensates}. In Fig. \ref{energy_glueball} the energy density is plotted. We see that for the existence of this regular solution it is absolutely necessary to have the different space distribution of two condesates $\phi^a(r)$ and $\phi^m(r)$. At the infinity $\phi^a \rightarrow \mu_1$ but $\phi^m \rightarrow 0$ that coincides with the condition \eqref{5-100}.
\par 
The similiraty between two approaches is that the ordered phase in CCM has the same essence as in SCSM. The difference is that two disordered phases $\phi^a$ and $\phi^m$ in CCM are constructed from two sets of gauge potential $A^a \in SU(2)$ and 
$A^m \in SU(3)/\biggl( SU(2) \times U(1) \biggl)$ but in SCSM two condensates are build from SU(2)/U(1) coset. In other words each off-diagonal $A^m_\mu (m=1,2)$ in SCSM is decomposed on spinless bosonic scalars $\psi_{1,2}$ and one color-neutral spin-one vector $\vec e_\mu$ but in CCM such decomposition corresponds to section \ref{momentum} where the ordered phase (color-neutral spin-one) vector is $A^8_\mu$ and disordered phases (spinless bosonic scalars) are $\phi^a$ and $\phi^m$ and obtained from SU(2) and 
$SU(3)/\biggl( SU(2) \times U(1) \biggl)$ components of SU(3) gauge potential correspondingly.
\par 
The advantage of CCM is that it allows us to calculate the functions $\phi^{a,m}(r)$ in contrast with SCSM which consider spatially uniform condensate only. 

\section{Conclusions}

In this letter we have compared two non-perturbative approahes in QCD: the collective coordinate method and the spin-charge separation. We have seen that both approaches have some close connection: the existence of two condensates which are necessary to confinement. The difference is that the first approach has the possibility to give us the space distribution of both condensates in contrast with the second approach which may give us spatially uniform distribution of these condensates only.

\end{document}